# Electronic-Structure Correlations Governing Superconductivity in Nb-Based High-Entropy Alloys

Md Sabbir Hossen Bijoy[1,*], Vladislav Korostelev[1,*], Deva Prasaad Neelakandan[1], Harshil Goyal[2], Steven E. Porterfield[1], Youming Xu[3], Shuchen Li[3], Xi Chen[3], Mark Adams[2], Barton C. Prorok[1], Konstantin Klyukin[1], Chanho Lee[1], Fariborz Kargar[1,†]

[1]Materials Research and Education Center, Department of Mechanical Engineering, Auburn University, Auburn, Alabama 36849 USA

[2]Department of Electrical and Computer Engineering, Auburn University, Auburn, Alabama 36849 USA

[3]Department of Electrical and Computer Engineering, University of California, Riverside, CA, 92521, USA

---

[*] These authors contributed equally.
[†] Corresponding author: fkargar@auburn.edu (F.K.)

# Abstract

Superconducting high-entropy alloys have recently emerged as a new platform for exploring superconductivity in highly disordered metallic systems and may offer advantages for applications requiring mechanical robustness and tolerance to extreme environments. Yet the mechanisms that govern their superconductivity – particularly the roles of lattice distortion and complex local order, both inherent to high-entropy alloys – remain unclear. The conventional valence-electron-concentration rule fails to reliably predict superconducting behavior, motivating a correlation analysis that links performance to electronic structure and lattice disorder. Here, we study a systematic series of niobium-based body-centered-cubic high-entropy alloys, from binary to quinary compositions, designed to investigate the electronic and structural effects and identify the dominant factors controlling superconductivity. Our experimental results reveal that the superconducting critical properties evolve non-monotonically with alloy complexity. Interestingly, alloys with greater lattice distortion can still achieve higher critical temperature and upper critical field. These observations are corroborated by first-principles and Eliashberg analyses, which identify the position of the niobium $d$-band relative to the Fermi level as the primary driver of electron–phonon coupling, critical temperature, and upper critical field, with lattice distortion serving as a secondary modifier that generally weakens coupling. We consolidate these findings into a detailed correlation map linking superconducting properties to electronic-structure fingerprints and vibrational signatures, establishing a mechanism-informed design strategy for superconducting high-entropy alloys with enhanced critical temperature and field.

## Introduction

High-entropy alloys (HEAs), also referred to as multi-principal-element alloys (MPEAs), represent a rapidly expanding class of materials in which several elements are mixed in near-equiatomic proportions within a single crystallographic lattice[1–7]. This compositional complexity produces strong local chemical disorder and lattice distortion, profoundly modifying electronic states and phonon spectra, and giving rise to exceptional mechanical properties such as high strength, radiation tolerance, and thermal stability[3,8–13]. Beyond structural applications, several HEAs have recently been found to exhibit superconductivity, opening opportunities for robust superconducting materials suitable for extreme environments[3,14–17]. The first reported superconducting HEA, $Ta_{0.34}Nb_{0.33}Hf_{0.08}Zr_{0.14}Ti_{0.11}$, is a type-II phonon-mediated superconductor with a transition temperature $T_c$~7.3 K and an upper critical magnetic field of $H_{c2}$~8 T[14]. Such high upper critical fields make HEA superconductors promising for applications requiring magnetic-field resilience, including MRI systems and radiation-tolerant superconducting components. However, the emergence of robust superconductivity in such chemically disordered systems is counterintuitive, as disorder is generally expected to disrupt electronic coherence and suppress superconducting pairing. This raises a fundamental question: whether superconductivity in HEAs is governed by conventional descriptors such as valence electron count (VEC), or by disorder-driven modifications to the electronic structure and phonon-mediated interactions. Resolving this question is essential for establishing predictive design principles for superconducting HEAs.

Although some currently reported HEA superconductors exhibit relatively modest critical temperatures as high as $T_c$~10 K, there is no fundamental limitation preventing the realization of higher-$T_c$ systems. Experimental studies indicate that key superconducting parameters – including critical current density, $j_c$, critical magnetic fields, $H_c$, and $T_c$ – can be tuned through compositional design[3,16,18]. In conventional body-centered-cubic transition-metal systems, such trends are often interpreted through empirical relationships such as the Matthias rule, which correlates $T_c$ with VEC[3,15,16,18–20]. However, the intrinsic disorder in HEAs alters both the electronic density of states near the Fermi level and the phonon spectrum governing electron–phonon coupling. As a result, superconductivity in HEAs cannot be understood solely through electron counting, and determining how electronic structure, phonon dynamics, and lattice distortion jointly govern $T_c$ and related critical parameters remains a central challenge.

In this work, we investigate these relationships using a systematic series of Nb-based high-entropy alloys spanning binary to quinary compositions while maintaining a body-centered-cubic crystal structure. By combining electrical transport, thermodynamic, and magnetic measurements with first-principles calculations and Eliashberg analysis, we disentangle the roles of electronic structure and lattice disorder in governing superconductivity. We find that superconducting properties evolve non-monotonically with increasing alloy complexity and lattice distortion. In particular, the first moment of the Nb $d$-band center relative to the Fermi level emerges as the primary electronic descriptor controlling electron–phonon coupling strength and the superconducting transition temperature, while lattice distortion acts as a secondary modifier that generally weakens coupling. By integrating experimental observations with theoretical calculations, we construct correlation maps linking superconducting parameters to key electronic and structural descriptors. Together, these results establish a mechanism-informed design framework for engineering HEA superconductors with enhanced transition temperatures and improved magnetic-field performance.

## Results

**Alloy Preparation and Characterization**

Elemental Nb is a prototypical conventional superconductor with strong electron–phonon coupling, and many of its alloys retain superconducting properties[14,19,21–27]. To systematically examine how chemical complexity and lattice distortion affect this behavior, we prepared a series of Nb-based multi-principal-element alloys, NbTa, NbTaTi, NbTaTiV, NbV, and NbTaTiVZr, spanning binary to quinary configurations. Figure 1a shows schematic BCC structures of representative NbTa and NbTaTiVZr alloys, highlighting local lattice distortions induced by atomic size mismatch, with the magnitude of distortion intentionally exaggerated for visual clarity. These compositions were selected to incrementally introduce atomic-size mismatch while maintaining a BCC crystal structure. The constituent elements belong to the 3$d$ (Ti, V), 4$d$ (Nb, Zr), and 5$d$ (Ta) transition-metal series, providing distinct atomic radii and electronic configurations, enabling systematic tuning of lattice strain and electronic character.

Rod-shaped samples were synthesized by vacuum arc-melting of high-purity elements (99.99 wt%) under argon, following procedures described in Methods. Phase purity and structural homogeneity of the samples were confirmed using high-energy synchrotron X-ray diffraction (XRD) (beamline 11-ID-C, Advanced Photon Source, Argonne National Laboratory, US). A representative diffraction pattern for NbTa is shown in Fig. 1b, with corresponding data for the other alloys provided in Supplementary Fig. 1. All samples exhibit diffraction patterns consistent with a single-phase BCC structure, with all major peaks accurately indexed to the corresponding crystallographic planes. To further verify phase purity and assess microstructural uniformity, electron backscatter diffraction (EBSD) mapping was performed on a representative alloy, NbTa, as shown in Fig. 1c. The uniform color contrast across each EBSD map confirms a single-phase BCC microstructure with no detectable secondary phases or compositional segregation. Additional scanning electron microscopy (SEM) images for all alloys are provided in Supplementary Fig. 2. The average lattice distortion, $\bar{u}^D$, was quantified from effective interatomic distances and alloy-induced changes in atomic volume, as detailed in the Methods section. As shown by the pink circles in Fig. 1d, $\bar{u}^D$ increases systematically across the alloy series – from NbTa through NbTaTi, NbTaTiV, and NbV to NbTaTiVZr – spanning 0.001 Å to 0.183 Å, consistent with the intended increase in atomic-size mismatch and chemical disorder. By contrast, VEC, defined as $VEC = \sum_i x_i v_i$, where $x_i$ and $v_i$ are the atomic fraction and valence electron count of the element $i$, respectively, does not follow the same trend as $\bar{u}^D$ (Fig. 1d, blue circles). VEC is commonly used to correlate with the superconducting transition temperature in HEAs[15,16,18]. This deliberate divergence between lattice distortion and VEC trends enables us to examine their respective roles in governing superconducting properties independently within a structurally uniform BCC framework. The elemental parameters used in calculating $\bar{u}^D$ and VEC are summarized in Supplementary Table 1.

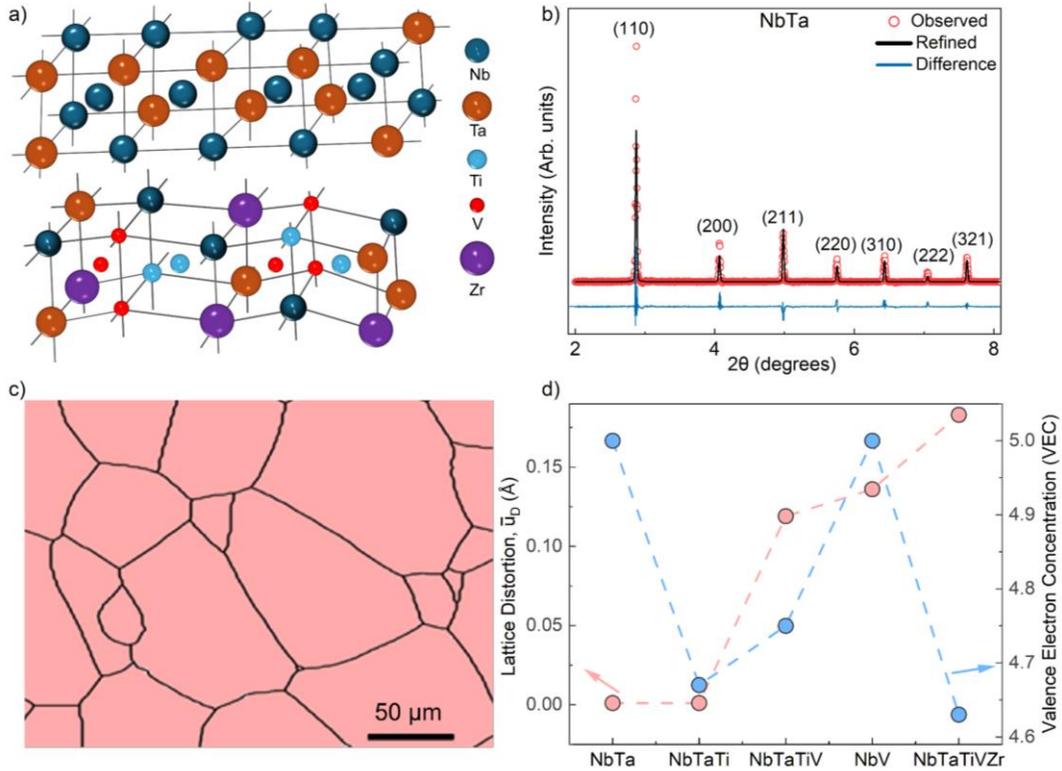

[**Figure 1 | Structural characterization and lattice distortion in Nb-based high entropy alloys.** (**a**) Schematic illustration of the body-centered cubic (BCC) crystal structure for representative Nb-based alloys of NbTa and NbTaTiVZr, highlighting the incorporation of multiple transition-metal elements and the resulting lattice distortion due to atomic-size mismatch. (**b**) Representative high-energy synchrotron X-ray diffraction (XRD) pattern for NbTa, showing the observed data (red circles), refinement (black curve), and difference profile (blue curve). All indexed reflections are consistent with a single BCC phase; corresponding XRD patterns for the remaining alloys are provided in Supplementary Fig. 1. (**c**) Electron backscatter diffraction (EBSD) map of NbTa, demonstrating a homogeneous single-phase BCC microstructure with no detectable secondary phases or compositional segregation. (**d**) Average lattice distortion, $\bar{u}^D$ (pink circles, left axis), and valence electron concentration (VEC) (blue circles, right axis) for the Nb-based alloy series. While $\bar{u}^D$ increases systematically, VEC does not follow the same trend, enabling independent assessment of lattice distortion and electronic effects within a structurally uniform BCC framework.]

**Superconducting Properties and Thermodynamic Signatures**

Following structural characterization, we examined the superconducting properties of the Nb-based alloys through temperature-dependent electric resistivity, specific heat, and magnetization measurements. These complementary techniques allow us to extract the critical parameters of superconductivity and correlate them with their electronic band structure, lattice distortion, and compositional complexity. The electric resistivity of all alloys was measured using the electric transport option (ETO) of the physical behavior measurement system (PPMS) (Quantum Design, USA). Figure 2a presents the temperature-dependent resistivity, $\rho(T)$, of the alloys measured during cooling from 15 K to 2 K under zero magnetic field. All samples exhibit metallic behavior in the normal state followed by a sharp drop to near-zero resistivity, confirming the emergence of superconductivity. The onset transition temperature, $T_c^{on}$, defined as the temperature where $\rho(T)$ first deviates from its normal-state value, and the transition width, $\Delta T_c = T_c^{on} - T_c^{zero}$, where $T_c^{zero}$ is the temperature at which $\rho(T) \rightarrow 0$, were extracted for each composition. The resulting ($T_c^{on}, \Delta T_c$) [K] pairs are (9.2, 0.2) for Nb, (6.0, 0.4) for NbTa, (3.5, 0.2) for NbV, (7.7, 0.2) for NbTaTi, (3.3, 0.4) for NbTaTiV, and (5.2, 0.4) for NbTaTiVZr. It is well established that structural or chemical disorder tends to broaden the superconducting transition in both conventional and unconventional systems[28,29]. The modest broadening of $\Delta T_c$ with increasing alloy complexity reflects enhanced chemical disorder and lattice distortion, which lead to spatial fluctuations in the local superconducting gap[30,31].

To investigate the thermodynamic transition and quantify related electron-phonon coupling parameters, we measured the specific heat, $c_p(T)$, of all samples using the thermal-relaxation method in the PPMS over the temperature range of 2 K to 300 K. The data are shown in Fig. 2b. As seen, all samples approach the Dulong–Petit limit of 3R at room temperature, where R is the universal gas constant[32]. Superconductivity manifests in $c_p(T)$ as a distinct jump at the transition temperature[29,32], as shown in Supplementary Fig. 3, where the data from Fig. 2b are replotted over a reduced temperature range of 2–15 K for clarity. It is common to plot $c_p/T$ as a function of $T^2$ to linearize the normal-state behavior, enabling clear separation of the electronic, $c_e$, and phononic, $c_{ph}$, specific heats and allowing $T_c$ to be quantitatively extracted from the discontinuity associated with the superconducting transition (Fig. 2c). The data in the normal metallic state region above $T_c$ were fitted using $c_p/T = \gamma + \beta T^2$, where the *y*-intercept, $\gamma$, and slope, $\beta$, of the linear fittings represent the electronic and phononic contributions to heat capacity, respectively. The fits yield $\gamma$ values in the range of 5.23–8.92 [mJmol$^{-1}$K$^{-2}$] and $\beta$ values of 0.072–0.178 [mJmol$^{-1}$K$^{-4}$]. The corresponding Debye temperatures, $\Theta_D = (12\pi^4 R/5\beta)^{1/3}$, are 259 K (Nb), 239 K (NbTa), 300 K (NbV), 222 K (NbTaTi), 266 K (NbTaTiV), and 250 K (NbTaTiVZr). The calculated $\Theta_D$ values for Nb and NbTa agree well with the literature[33]. In general, increasing alloy complexity and average molar mass, $\bar{M}$, lead to lower $\Theta_D$, consistent with mass-dependent lattice softening and reduced phonon frequencies in multicomponent systems[18,34–36]. Minor deviations from the monotonic behavior of $\Theta_D$ vs $\bar{M}$ arise from variations in bonding strength, lattice distortion, and local force constants among the alloys.

To further interpret these thermodynamic results within the framework of Bardeen–Cooper–Schrieffer (BCS) theory, we examine the relation between $T_c$, $\Theta_D$, and electron–phonon coupling, $\lambda_{e-ph}$. According to the BCS theory, the $T_c$ scales proportionally with $\Theta_D$ as $T_c \sim \Theta_D \exp(-1/(\lambda_{e-ph} - \mu^*))$, where $\mu^*$ is the Coulomb pseudopotential accounting for screened electron–electron repulsion[37,38]. While this relation implies that higher $\Theta_D$ should yield higher $T_c$ for comparable coupling strength, the Nb-based HEAs deviate from this simple dependence. This deviation indicates that changes in $\lambda_{e-ph}$ and in the electronic density of states across different alloy systems, play a central role in determining $T_c$. Experimental values of $\lambda_{e-ph}$ for each alloy were estimated by inverting the Allen–Dynes modified McMillan relation (Supplementary Equation S1) using the measured $T_c$ and $\Theta_D$, assuming $\mu^* = 0.13$[18,39]. The resulting $\lambda_{e-ph}$ values are 0.90 (Nb), 0.78 (NbTa,), 0.62 (NbV), 0.91 (NbTaTi), 0.63 (NbTaTiV), and 0.75 (NbTaTiVZr), indicating stronger coupling in NbTaTi and near-BCS weak coupling in the high-distortion pair NbV and NbTaTiV. These variations in coupling strength point to significant variation in electronic properties of the alloys that extend beyond lattice softening, as examined in the following section.

The stronger $\lambda_{e-ph}$ inferred from the Allen–Dynes analysis is also reflected in the normalized specific-heat jump, $\Delta c_e/\gamma T_C$, which separates the alloys by their superconducting transition strength[18,40]. The electronic specific heat, $c_e$, was obtained by subtracting the phononic part, $c_{ph} = \beta T^3$, from the total heat capacity, and plotted as a function of temperature in Fig. 2d. As seen, NbTaTi exhibits the largest normalized jump of 2.42, while Nb, NbV, NbTa, and NbTaTiV show values close to the weak-coupling BCS limit of 1.43, and NbTaTiVZr shows an intermediate value of 1.78. The electronic specific heat below $T_c$ of all alloys follows an exponential decay, $c_e \sim exp(-\Delta_0/k_B T)$, characteristic of a fully gapped superconducting state[37]. Fits to the data yield superconducting gap values, $\Delta_0$, of 1.62 meV (Nb), 0.92 meV (NbTa), 1.44 meV (NbTaTi), 0.51 meV (NbTaTiV), 0.56 meV (NbV), and 0.84 meV (NbTaTiVZr). Among all compositions, NbTaTi displays the largest $\Delta c_e/\gamma T_C$ and $\Delta_0$, consistent with its strongest electron–phonon coupling and highest $T_c$. The combined thermodynamic and electronic analyses confirm that superconductivity in Nb-based HEAs evolves nonmonotonically with alloy complexity. The dimensionless gap ratio, $2\Delta_0/k_B T_C$, ranges from 3.54 to 4.20, with NbTaTi exhibiting a clear deviation from the weak-coupling BCS value of 3.53, indicative of enhanced electron–phonon coupling, while the remaining alloys remain close to the BCS limit. The extracted thermodynamic parameters, including $\gamma$, $\beta$, $\Theta_D$, $\Delta c_e/\gamma T_C$, and $2\Delta_0/k_B T_C$, are summarized in Supplementary Table 2.

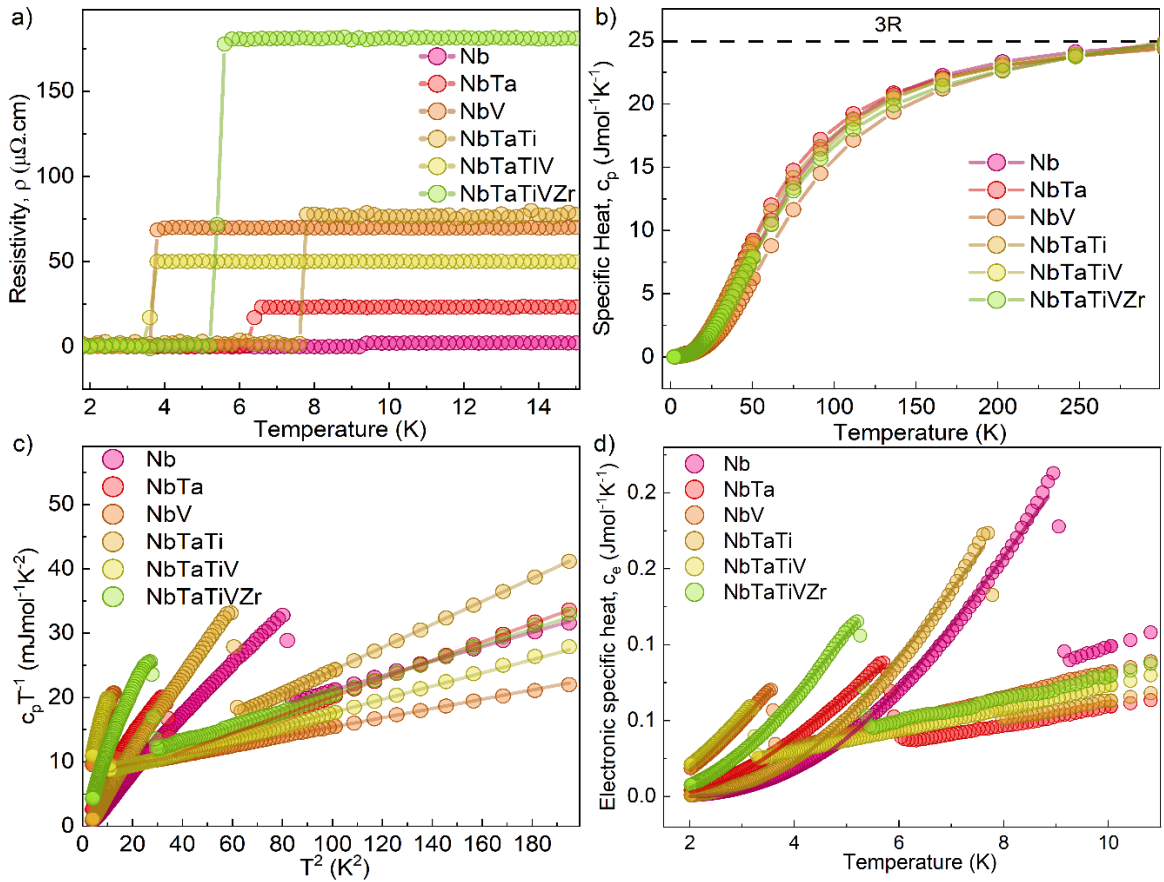

[**Figure 2 | Temperature-dependent electrical resistivity and specific-heat characteristics of Nb-based high entropy alloys.** (**a**) Temperature-dependent electrical resistivity of alloys, showing sharp superconducting transitions characterized by an abrupt drop to near-zero resistivity below $T_c$. (**b**) Specific heat, $c_p$, of the examined alloys in the temperature range of 2 K to 300 K, showing convergence toward the Dulong–Petit limit at high temperatures. $R$ is the universal gas constant. (**c**) $c_p/T$ plotted as a function of $T^2$, exhibiting linear dependence in the normal state, $T > T_c$, due to electronic and phononic contributions. The slope and *y*-intercept of the linear fits

yield the Sommerfeld electronic coefficient, γ, and phonon contribution β. Below $T_c$, *the data shows an exponential decay, indicative of a fully gapped superconducting state.* **(d)** Isolation of electronic specific heat, $c_e$ to quantify the strength of electron-phonon coupling and superconducting gap. The normalized jump at transition, $\Delta c_e/\gamma T$, reflects the electron-phonon coupling strength, while the exponential decay at low temperatures quantifies the superconducting energy gap.]

We next extract the critical magnetic fields by conducting magnetization measurements using the vibrating-sample magnetometer (VSM) option of the PPMS. Figure 3a presents the temperature-dependent magnetization, $M(T)$, measured under zero-field-cooled (ZFC) and field-cooled (FC) conditions for representative alloys of NbTaTi and NbTaTiV in an applied field of $H{\sim}50$ Oe upon cooling the samples from 15 K to 2 K. The data for other samples are provided in Supplementary Fig. 4. All alloys exhibit a clear diamagnetic transition, with ZFC and FC curves separating near their respective $T_c$. The separation of the ZFC and FC curves reflects magnetic flux pinning, a characteristic of Type-II superconductors, while the sharp diamagnetic response below $T_c$ confirms bulk superconductivity[41,42]. The transition temperatures obtained from $M(T)$ agree closely with those extracted from resistivity and specific-heat measurements, further confirming the bulk nature of superconductivity in all samples.

Following the temperature-dependent magnetization measurements, a series of isothermal $M(H)$ measurements were carried out under varying external magnetic fields to determine the lower critical field, $H_{c1}$, for each alloy. The temperature dependence $H_{c1}(T)$ was extracted from the first deviation of the low-field $M(H)$ curves from linearity, marking the onset of magnetic flux penetration into the sample (Supplementary Fig. 5). The resulting $H_{c1}(T)$ data were fitted using the Ginzburg–Landau (GL) relation $H_{c1}(T) = H_{c1}(0)(1 - (T/T_c)^2)$, where $H_{c1}(0)$ is the extrapolated zero-temperature value[29]. The corresponding GL fits, shown in Fig. 3b, yield $H_{c1}(0)$ values of 29.6 mT (NbTa), 20.0 mT (NbTaTi), 10.6 mT (NbV), 12.8 mT (NbTaTiV), and 13.1 mT (NbTaTiVZr). Alloys with smaller lattice distortion (NbTa, NbTaTi) exhibit higher $H_{c1}$, consistent with smaller magnetic penetration depths, $\lambda_0$, whereas compositions with greater lattice distortion (NbV, NbTaTiV, and NbTaTiVZr) show reduced $H_{c1}$ values.

The upper critical field, $H_{c2}$, was estimated from the suppression of the diamagnetic response with increasing applied magnetic field, where the diamagnetic signal is suppressed toward zero, signifying the complete destruction of superconductivity[14,37] (see Supplementary Fig. 6). Figure 3c shows $H_{c2}$ as a function of temperature for each alloy. The consistency of $T_c$ determined from specific heat, electrical resistivity, and magnetic susceptibility, together with the vanishing electronic specific heat, i.e., $c_{el}/T \to 0$ at low temperature, demonstrates homogeneous bulk superconductivity. The data were fitted using the GL form, $H_{c2}(T) = H_{c2}(0)[(1 - (T/T_c)^2)/(1 + (T/T_c)^2)]$[29]. The extracted $H_{c2}(0)$ values are 0.98 T (NbTa), 2.43 T (NbV), 9.94 T (NbTaTi), 1.68 T (NbTaTiV), and 7.7 T (NbTaTiVZr). The coherence length, $\xi_0$, for each sample was calculated according to $\xi_0 = (\phi_0/2\pi\mu_0 H_{c2}(0))^{1/2}$, where $\phi_0 = 2.07 \times 10^{-15}$ [Wb] is the flux quantum. The penetration length, $\lambda_0$, was calculated iteratively from $H_{c1}(0) = (\phi_0/4\pi\lambda_0^2)[\ln \kappa + 0.5]$. Here, $\kappa_{GL} = \lambda_0/\xi_0$ is the Ginzburg-Landau parameter. For all alloys, we obtained $\kappa_{GL} > 1/\sqrt{2}$, confirming that they exhibit type-II superconductivity[29,32]. Figure 3d summarizes the $T_c$ and $H_{c2}$, with alloys arranged along the $x$-axis in order of increasing lattice distortion. Both $T_c$ and $H_{c2}$ peak for NbTaTi and, to a lesser extent, NbTaTiVZr, demonstrating a nonmonotonic evolution of superconductivity with lattice distortion and indicating that factors beyond simple VEC scaling govern the superconducting properties of these alloys. All critical and related parameters extracted from electronic, heat capacity, and magnetic measurements are summarized in Table 1.

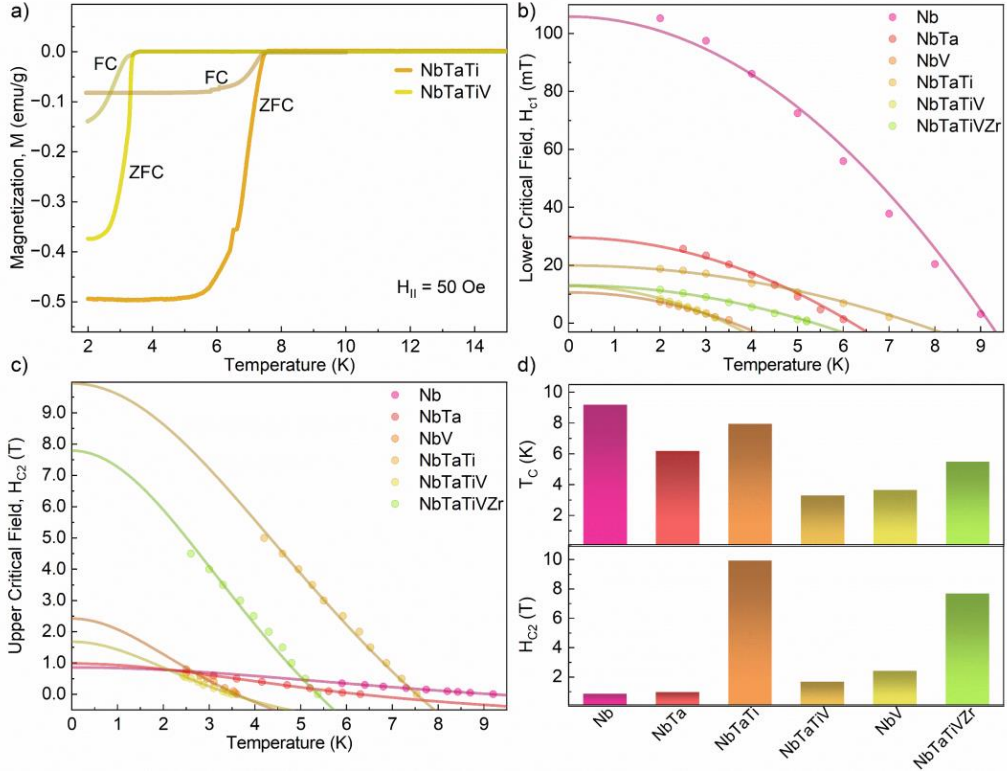

**[Figure 3 | Magnetic characterization and critical-field behavior of Nb-based high-entropy alloys.** **(a)** Temperature-dependent magnetization for two representative alloys, NbTaTi and NbTaTiV, measured under zero-field-cooled (ZFC) and field-cooled (FC) conditions at an external magnetic field of $H = 50$ Oe. The onset of diamagnetism marks the superconducting transition, while the deviation between FC and ZFC curves below $T_c$ reflects the strength of flux pinning in the superconducting state, confirming their Type-II superconductivity. **(b, c)** Lower and upper critical fields, $H_{c1}$ and $H_{c2}$, of the alloys as a function of temperature, fitted using the Ginzburg–Landau (GL) models. **(d)** Critical temperatures, $T_c$, and upper critical field, $H_{c2}$ extracted from magnetic susceptibility measurements. Elements in the *x*-axis are arranged in order of increasing lattice distortion.]

[Table 1 | Summary of critical and other related parameters of the Nb-based high-entropy alloys]

| Sample | $\bar{u}^D$ (Å) | $T_c$ (K) | $H_{c1}$ (mT) | $H_{c2}$ (T) | $\Theta_D$ (K) | $\lambda_{e-ph}$ | $\Delta_0$ (meV) | $\xi_0$ (nm) | $\lambda_0$ (nm) | $\kappa_{GL}$ |
|---|---|---|---|---|---|---|---|---|---|---|
| Nb | - | 9.2 | 105.9 | 0.86 | 259 | 0.90 | 1.62 | 19.6 | 45.8 | 2.3 |
| NbTa | 0.001 | 6.0 | 29.6 | 0.98 | 239 | 0.78 | 0.92 | 18.3 | 113.8 | 6.2 |
| NbTaTi | 0.001 | 7.7 | 20.0 | 9.94 | 222 | 0.91 | 1.44 | 5.8 | 180.1 | 31.1 |
| NbTaTiV | 0.119 | 3.3 | 12.8 | 1.68 | 266 | 0.63 | 0.51 | 14.0 | 201.8 | 14.4 |
| NbV | 0.136 | 3.5 | 10.6 | 2.43 | 300 | 0.62 | 0.56 | 11.7 | 232.9 | 19.9 |
| NbTaTiVZr | 0.183 | 5.2 | 13.1 | 7.7 | 250 | 0.75 | 0.84 | 6.5 | 225.6 | 34.7 |

## Discussion

Lattice distortion is an inherent characteristic of high-entropy alloys, arising from the random distribution of elements with different atomic sizes and bonding environments[7–9,43]. In our Nb-based systems, this distortion varies with composition and strongly influence superconductivity. The experimental data demonstrate that $T_c$, $\lambda_{e-ph}$, and $H_{c2}$ evolve nonmonotonically with both VEC and mean bond-length distortion, $\bar{u}^D$. Such behavior cannot be fully captured by Matthias' rule[44], which correlates $T_c$ primarily with VEC and neglects the effects of lattice distortion. To disentangle these intertwined structural and electronic effects, we combined density functional theory (DFT) calculations with Eliashberg and Allen–Dynes formalisms[39,45], to quantify how atomic-scale disorder modulates electronic and phononic band structures, electron–phonon coupling, and ultimately superconductivity in Nb-based HEAs.

To examine how local lattice distortion influences superconductivity, we first constructed low-distortion (LD) and high-distortion (HD) configurations for the alloys and analyzed their electronic and vibrational properties. The LD configuration corresponds to a special quasi-random structure (SQS) that preserves chemical randomness while keeping atoms close to their ideal BCC positions[46,47]. The HD configuration is selected from randomly generated atomic assignments that exhibit a broader distribution of nearest-neighbor bond lengths, corresponding to a higher bond-length variance. These configurations represent limiting cases of local atomic arrangements that bracket the range of lattice distortions expected in chemically disordered HEAs. Figure 4 summarizes the resulting atomistic picture and organizes the findings in a cause-and-effect sequence linking atomic disorder, electronic structure, and superconducting properties.

Figure 4a shows the bond-length distribution for a representative alloy, NbTaTiV, in both LD and HD configurations. The *x*-axis represents the fractional deviation of nearest-neighbor bond lengths from the ideal BCC nearest-neighbor distance of $(\sqrt{3}/2)a$, where a is the lattice parameter for each fully optimized specific alloy composition and configuration. Negative values correspond to compressed bonds, while positive values indicate stretched ones. The *y*-axis represents the probability density of bond-length deviations. Higher values correspond to a greater likelihood of finding bonds with a given fractional deviation, $\mu$, from the ideal BCC distance. The resulting histograms were fitted with Gaussian functions to extract the mean deviation, μ, which quantifies the degree of lattice distortion in each configuration. The LD structure exhibits a Gaussian centered at $\mu \sim -0.20\ \%$, while the HD structure displays a left-shifted Gaussian centered at $\mu \sim -1.16\ \%$.

This shift toward shorter bonds reflects stronger local strain and is accompanied by a reduction in the calculated superconducting transition temperature from 8.7 K to 5.3 K. These results demonstrate that even at fixed stoichiometry, local atomic configurations alone can significantly modify $T_c$, highlighting the intrinsic sensitivity of superconductivity to local lattice distortion.

To understand how lattice distortion influences superconductivity, we next examined how atomic disorder modifies the electronic band structure, which directly governs the density of states, DOS, at the Fermi level and, in turn, the electron–phonon coupling strength. Figure 4b shows the calculated band structure and projected density of states (PDOS) for a representative alloy, NbTaTiV, in its LD configuration. Among the alloying elements, Nb exhibits the highest $T_c$ due to its strong electron-phonon coupling. Accordingly, Nb-derived $d$-states couple more strongly to lattice vibrations and therefore make the dominant contribution to superconductivity, even though other elements such as Ti and V display larger projected DOS near the Fermi level in the alloy (Fig. 4b). This motivates the identification of an electronic descriptor that captures how the Nb $d$-states are positioned relative to the Fermi level.

In conventional superconductors, pairing is governed by electronic states near $E_F$, both through their density and their energetic proximity to the Fermi level. Thus, a descriptor that quantifies the placement of the Nb $d$-manifold relative to $E_F$ is more physically meaningful than one that simply counts electrons, such as VEC. Note that standard descriptors such as VEC or the total $d$-band filling, $N_{tot}^d$ do not contain this energetic information. As shown in Supplementary Fig. 7, VEC correlates linearly with $N_{tot}^d$, but both descriptors fail to distinguish electronic structures that share the same stoichiometry yet differ only by local disorder represented by LD and HD configurations. A meaningful electronic quantity that captures this behavior is the first moment of the occupied Nb $d$-band, $\mu_d^{Nb}$, defined as, $\mu_d^{Nb} = \frac{\int_{-\infty}^{E_F} E\, \text{DOS}_{\text{Nb-}d}(E)\, dE}{\int_{-\infty}^{E_F} \text{DOS}_{\text{Nb-}d}(E)\, dE}$. A summary of all symbols and definitions used throughout the manuscript is provided in the Supplementary Information (Section 1, Nomenclature). $\mu_d^{Nb}$ represents the energy centroid of the occupied Nb $d$-projected DOS relative to the Fermi level (Methods; details in Supplementary Information Section 5.2). By isolating only the *occupied* Nb $d$-states, i.e., the states that contribute directly to electronic transport at $E_F$, $\mu_d^{Nb}$ provides a direct, physically grounded measure of how lattice distortion shifts the Nb $d$-manifold, distinguishing LD and HD configurations in ways that VEC and $N_{tot}^d$ cannot. Figure 4c shows the impact of lattice distortion and chemical alloying on displacing $\mu_d^{Nb}$ with respect to the Fermi level (black dashed line). The results are exhibited for two representative alloys of NbV and NbTaTiV under both LD and HD configurations. The vertical dashed lines mark the extracted values of $\mu_d^{Nb}$ for each composition and configuration. As seen, alloying elements such as Ta and Ti tend to shift $\mu_d^{Nb}$ upward by ~0.2 eV relative to NbV, increasing $N(E_F)$ and strengthening electron–phonon coupling. These results demonstrate that an upward shift of $\mu_d^{Nb}$ toward $E_F$ enhances the electronic density of states and strengthens electron–phonon coupling. At fixed composition, LD configurations place $\mu_d^{Nb}$ closer to $E_F$, whereas HD configurations push it to lower energies, an electronic signature of lattice distortion that reduces the available $d$-electron spectral weight at the Fermi level and weakens coupling . These results highlight the strong sensitivity of $\mu_d^{Nb}$ to lattice distortion and alloying, which is later used to develop the correlation maps among key parameters governing superconducting critical properties in HEAs.

While the position of $\mu_d^{Nb}$ defines how readily Nb $d$-electrons participate in Cooper pairing, the phonon spectrum determines the lattice modes that mediate this pairing. To connect these two aspects and to understand the impact of distortion on phonon states, we quantify the electron–phonon interaction directly using the Eliashberg spectral function, $\alpha^2 F(\omega)$[48,49]. In this framework, $\omega$ is the phonon angular frequency, $F(\omega)$ represents the phonon density of states, while $\alpha^2$ denotes the squared electron–phonon matrix element averaged over the Fermi surface. A larger $\alpha^2 F(\omega)$ value indicates that the corresponding vibrational modes are highly effective mediating Cooper pairing[38,48,50]. Within the Allen–Dynes modification of the McMillan equation, the superconducting transition temperature is given by[39]

$$T_c = (W(\omega)/1.2) \exp[-1.04(1 + \lambda_{e-ph})/(\lambda_{e-ph} - \mu^*(1 + 0.62\lambda_{e-ph}))],$$

where $W(\omega)$ is the logarithmic average phonon frequency representing, the effective phonon energy scale weighted by electron–phonon coupling. Both $W(\omega)$ and $\lambda_{e-ph}$ are directly derived from $\alpha^2F(\omega)$ and can be calculated from Eliashberg spectral function through frequency-dependent integrals of $\alpha^2F(\omega)/\omega$ (see Methods; Supplementary Information Section 6.2). Figure 4d shows $\alpha^2F(\omega)$ and the corresponding cumulative $\lambda_{e-ph}$ as a function of phonon angular frequency, $\omega$, for representative alloys. The spectral profiles are determined primarily by chemical composition, with additional alterations from lattice distortion. In terms of composition, the lighter NbV alloy concentrates spectral weight at higher phonon frequencies and therefore exhibits a larger $W(\omega) \sim 17.0$ meV compared with the heavier NbTaTiV alloy, whose dominant coupled modes occur near $W(\omega) \sim 14.6$ meV. Lattice distortion further modulates the phonon spectrum by altering local atomic stiffness, though no clear correlation emerges between distortion and $W(\omega)$. As seen, for NbV, the LD configuration shows a slightly higher $W(\omega)$ than the HD configuration, whereas for NbTaTiV the opposite trend is observed. However, for both compositions, the LD configurations consistently exhibit larger $\lambda_{e-ph}$ values than the HD configurations. These results indicate that variations in $T_c$ are governed primarily by changes in $\lambda_{e-ph}$, rather than systematic trends in $W(\omega)$.

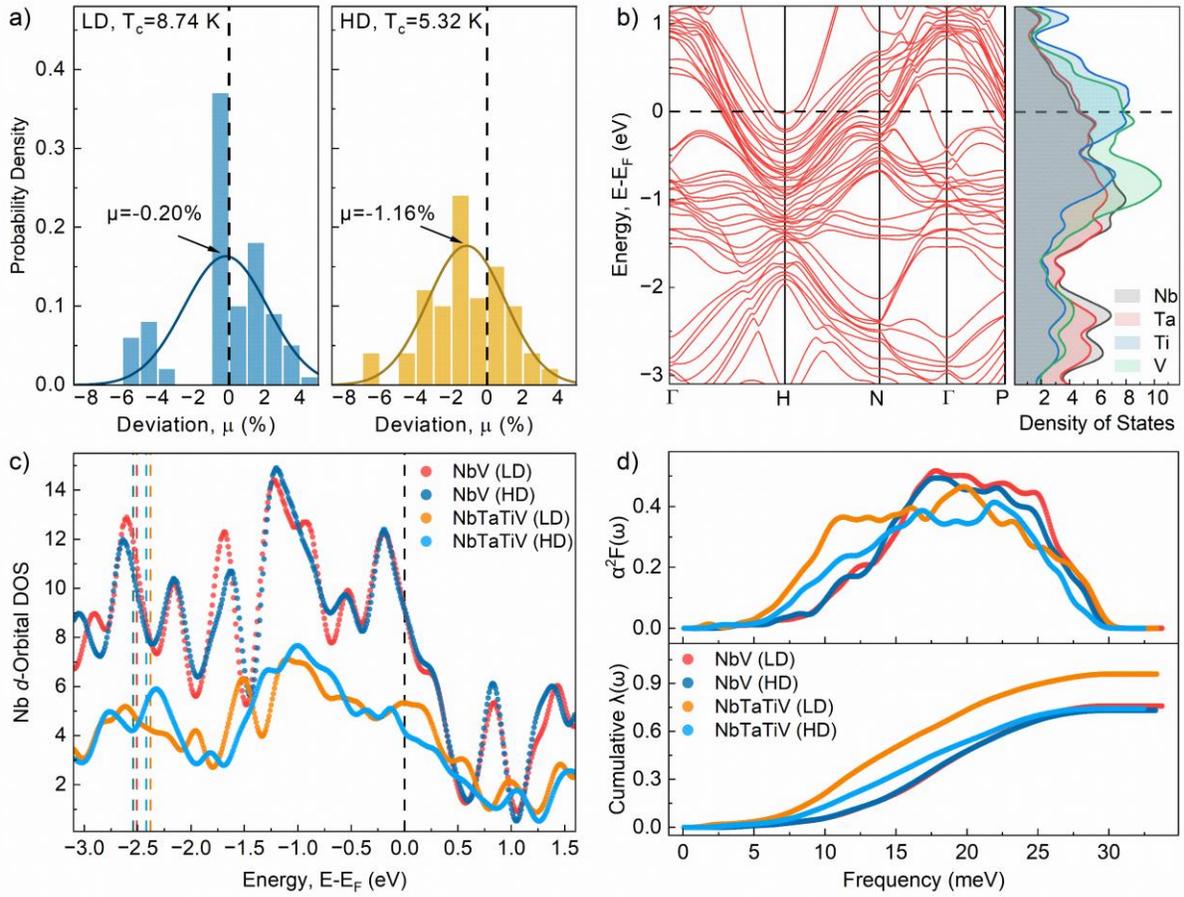

**[Figure 4 | Calculated structural, electronic, and superconducting properties of Nb-based high-entropy alloys. (a)** Statistical distributions of nearest-neighbor distances for NbTaTiV in low-distortion (LD) and high-distortion (HD) configurations. The dashed line denotes the ideal BCC distance (0 % deviation). LD and HD configurations show average bond-length deviations, $\mu$, of −0.20 % and −1.16 %, respectively, with calculated superconducting transition temperatures decreasing from 8.74 k to 5.32 K. The latter links enhanced lattice distortion within the same composition to weakened electron–phonon coupling and reduced $T_c$. **(b)** Representative electronic band structure for NbTaTiV. The Nb-derived bands that cross the Fermi level in the −1 eV to 0 eV range are characteristic of all Nb-based HEAs and form the electronic states responsible for superconductivity. **(c)** Nb-projected $d$-orbital density of states, $d$-DOS, for two alloys, NbV, and NbTaTiV. For the latter two, the data are presented for both LD and HD configurations. Vertical dashed lines mark the first moment of Nb d-band filling ($\mu_d^{Nb}$), highlighting their sensitivity to both chemical composition and local lattice disorder. **(d)** Eliashberg spectral functions $\alpha^2F(\omega)$ (upper panel), and cumulative electron–phonon coupling constants, $\lambda$ (lower panel), for the same set of alloys shown in (c). The $\lambda$ values are affected by both composition and the degree of lattice distortion, as exemplified by the LD and HD within the same alloy composition.]

The results presented above demonstrate that superconductivity in Nb-based HEAs arises from a delicate interplay among electronic structure, phonon dynamics, and local lattice disorder. This
Page | 17

complexity necessitates a correlation-based analysis to disentangle their coupled effects and identify the parameters that most strongly govern $T_c$ and other superconducting properties. Within the Allen–Dynes modified McMillan equation, $T_c$ depends explicitly on $W(\omega)$ and $\lambda_{e-ph}$, motivating us to examine how these two quantities evolve with $\mu_d^{Nb}$ and $\bar{u}^D$. Before presenting the full correlation maps, we first establish the pairwise relationships among these key descriptors to reveal the trends that ultimately control $T_c$. Note that in addition to the experimentally investigated alloys, we also included NbMo and NbTi as additional DFT reference compositions to extend the analysis beyond the measured alloy systems. NbMo, in particular, serves as an important special case because its $\mu_d^{Nb}$ lies far below the Fermi level while exhibiting minimal lattice distortion. This combination provides a valuable reference point for disentangling the dominant and secondary effects of $\mu_d^{Nb}$ and lattice distortion, $\bar{u}_d$, on $\lambda_{e-ph}$, $W(\omega)$, and $T_c$. We note that when the calculated logarithmic average phonon frequency, $W_\omega^{DFT}$ and $\lambda_{e-ph}^{DFT}$ are combined within the Allen–Dynes equation, the calculated critical temperature, $T_c^{DFT}$, closely matches experimental $T_c$ values (see Supplementary Fig. 9 and Supplementary Table 4). Because of this close agreement, the correlation analysis that follows uses $T_c^{DFT}$ rather than the experimental $T_c$, allowing us to evaluate trends directly within a self-consistent theoretical framework and to incorporate additional alloys that were not experimentally characterized in this work.

Figures 5a–c show the correlation between $W_\omega^{DFT}$, $\lambda_{e-ph}^{DFT}$, and $T_c^{DFT}$ with $\mu_d^{Nb}$, demonstrating that $\mu_d^{Nb}$ serves as a robust electronic descriptor for superconductivity. A strong negative correlation (R = –0.99) is found between $\mu_{Nb}^d$ and $W_\omega^{DFT}$, indicating that as the Nb's $d$-band center approaches the Fermi level, the lattice softens. This trend is consistent with the conventional electron–phonon coupling framework, in which an upward shift of the Nb $d$-band center toward $E_F$ tends to increase the electronic density of states at the Fermi level, thereby enhancing the phase space available for electron–phonon scattering. At the same time, the electron–phonon coupling constant $\lambda_{e-ph}^{DFT}$ increases nearly linearly with $\mu_d^{Nb}$ with R = 0.90, leading to a higher $T_c^{DFT}$. Within BCS and Eliashberg formalisms, $\lambda_{e-ph} \sim N(E_F)\langle I^2\rangle/M\langle\omega^2\rangle$, where $N(E_F)$ is the electronic density of states at the Fermi level and $\langle I^2\rangle$ is the electron–phonon matrix element. The observed increase in $\lambda_{e-ph}$ with $\mu_d^{Nb}$ is therefore consistent with an increase in $N(E_F)$ associated with the upward shift of the Nb $d$-band center. Consequently, a strong positive correlation between the $T_c^{DFT}$ and $\mu_d^{Nb}$ with R= 0.79 is obtained. Consistent with the established trends, NbMo follows the same correlations. Its low $\mu_d^{Nb}$, lying far below the Fermi level, results in the smallest $\lambda_{e-ph}^{DFT}$, yielding the lowest $T_c^{DFT}$. By comparison, the calculated average $d$-band center, $\mu_d^{tot}$, across all alloying elements exhibits only moderate correlations with $T_c^{DFT}$ (R = 0.44) and $\lambda_{e-ph}^{DFT}$ (R = 0.49) (Supplementary Fig. 10). These results highlight that superconductivity in these alloys is governed primarily by the Nb $d$-electronic states rather than the average $d$-electron environment. To place $\mu_d^{Nb}$ in context with commonly used descriptors, we next examined the performance of the VEC, which is frequently employed as a simplified predictor in HEA superconductivity studies. Unlike $\mu_d^{Nb}$, VEC lacks energetic information and therefore cannot distinguish electronic structures that differ only due to local disorder. Consistent with this limitation, VEC correlates only weakly with $T_c^{DFT}$ (R = –0.55) and moderately with $\lambda_{e-ph}^{DFT}$ (R = –0.73), while showing a strong correlation with $W_\omega^{DFT}$ (R = 0.93) (Supplementary Fig. 10). This reflects that higher VEC values generally stiffen phonon modes, raising $W_\omega$, but do not reliably capture electron–phonon coupling strength. Moreover, VEC does not capture the effects of local atomic order. For instance, LD and HD configurations of NbTaTiV have identical VEC values yet exhibit different $T_c^{DFT}$ values. These observations indicate that while VEC can serve as a coarse empirical metric, it cannot account for disorder-driven electronic and vibrational effects that strongly influence superconductivity [16,18,43,44].

Figure 5d-f correlates $W_\omega^{DFT}$, $\lambda_{e-ph}^{DFT}$, and $T_c^{DFT}$ with the mean bond-length distortion, $\bar{u}^D$. Among all alloys, NbMo (red circle) behaves as a clear outlier. Despite its minimal lattice distortion, NbMo displays an unusually low $T_c \approx 0.5$ K and $\lambda_{e-ph} \approx 0.16$ due to its deep $\mu_d^{Nb}$, high VEC, and strong Nb–Mo bonding, which localize electrons well below $E_F$ and suppress electron–phonon interactions[51–54]. Because NbMo deviates from the general trends, we explicitly assess its impact on the correlations in Figs. 5d–f by performing linear regressions both including (pink solid line) and excluding (blue dashed line) this outlier. When NbMo is included, increasing distortion overall weakens superconductivity. Among the three quantities, $W_\omega^{DFT}$ shows only a very weak dependence on $\bar{u}^D$, whereas alloys with larger $\bar{u}^D$ exhibit lower $\lambda_{e-ph}^{DFT}$ (R = +0.20) and reduced $T_c^{DFT}$ (R = +0.35). We note that $\bar{u}^D$ is negative for compressed, highly distorted structures and approaches zero as the distortion decreases; therefore, a positive correlation coefficient (R > 0) indicates that $W_\omega^{DFT}$, $\lambda_{e-ph}^{DFT}$, and $T_c^{DFT}$ increases as $\bar{u}^D$ moves toward zero. In other words, stronger distortion, *i.e.*, more negative $\bar{u}^D$ suppresses electron–phonon coupling and decreases $T_C^{DFT}$, even though the numerical trend appears as a positive slope. Excluding NbMo reveals the intrinsic distortion trend in the examined alloys in this study more clearly. $W_\omega^{DFT}$ remains essentially uncorrelated with $\bar{u}^D$, confirming that phonon frequencies are largely insensitive to lattice disorder. In contrast, the correlation between $\lambda_{e-ph}^{DFT}$ and $\bar{u}^D$ strengthens substantially – from R = 0.20 (including NbMo) to R = 0.74 (excluding NbMo) – demonstrating that lattice distortion significantly influences $\lambda_{e-ph}^{DFT}$ and, therefore $T_c$ for the superconducting alloys. This reduction in $\lambda_{e-ph}$ can be attributed to disorder-induced broadening and shifting of the electronic states, which reduces the effective spectral weight of Nb *d*-states at the Fermi level and weakens the electron–phonon interaction. The latter result implies that in superconducting alloys and in regimes where electronic descriptors such as $\mu_d^{Nb}$ and VEC are nearly identical, as in the LD and HD configurations of NbV, the extent of lattice distortion dictates the strength of electron–phonon coupling and ultimately sets the superconducting transition temperature in HEAs.

While the preceding analysis focused primarily on identifying the parameters governing the superconducting transition temperature, the present analysis extends this framework to include the broader set of quantities that collectively determine superconducting performance. Figure 5g presents the complete correlation matrix encompassing $T_c$, $H_{c1}$, and the upper critical field, $H_{c2}$, together with electronic, vibrational, and structural descriptors. In Fig. 5g, the color scale represents the sign and magnitude of the Pearson correlation coefficient, *R*, ranging from -1 to +1, with blue and red indicating positive and negative correlations, respectively. The circle size is proportional to the absolute value of the correlation coefficient, highlighting the relative strength of each pairwise relationship (see Supplementary Fig. 10 for exact *R* values). Among these, $H_{c2}$, which defines the operational magnetic-field limit for superconducting applications, exhibits a strong positive correlation with the $\mu_d^{Nb}$ and a secondary dependence on the average Nb atoms phonon frequency, $\bar{\omega}_{Nb}$. This parameter is calculated as the weighted average phonon band center obtained from the phonon density of states projected onto Nb atoms (See Supplementary information Section 6.1 for details). Alloys in which $\mu_d^{Nb}$ lies closer to $E_F$ while maintaining moderately softened phonon spectra sustain higher critical fields (see Supplementary Fig. 8b). This dual dependence demonstrates that the same electronic and lattice parameters that optimize $T_c$ also govern magnetic-field robustness, establishing a unified materials design strategy in which tuning the Nb $d$-band center through alloying or strain, while controlling lattice stiffness and disorder, enables simultaneous optimization of $T_c$, $\lambda_{e-ph}$, and $H_{c2}$.

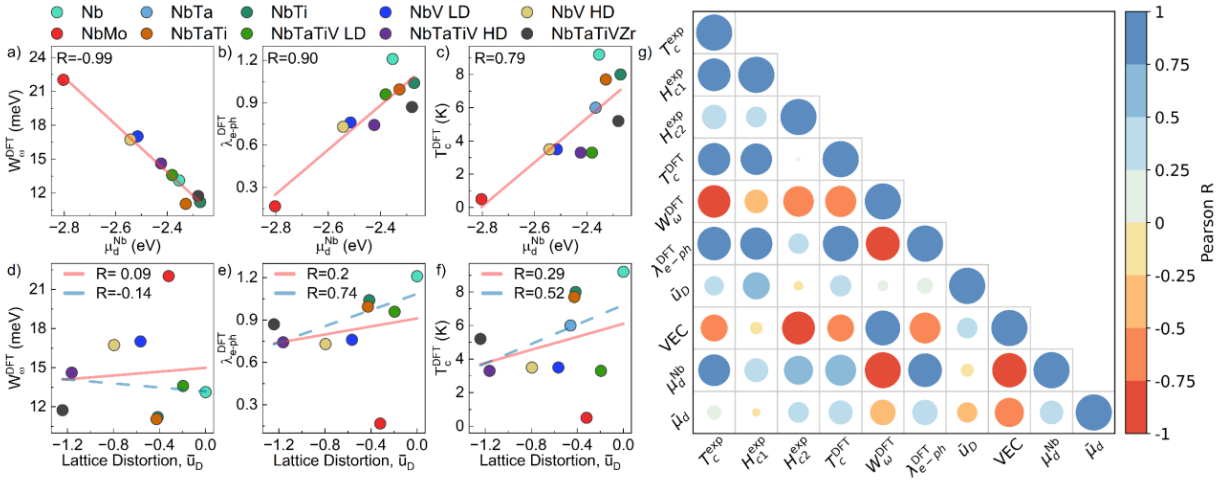

**Figure 5 |** Correlation matrix showing the relationships among different parameters influencing the superconducting properties of high-entropy alloys. The color scale represents the sign and magnitude of the Pearson correlation coefficient, while the bubble size denotes the strength of the correlation. Strong positive correlations (blue) indicate parameters that vary proportionally, whereas strong negative correlations (red) highlight parameters that exhibit opposing trends.

Our combined experimental and theoretical analysis establishes that superconductivity in Nb-based high-entropy alloys is governed by a hierarchy of electronic and structural parameters and cannot be reliably predicted by valence electron concentration alone. Across a systematic series of alloys, we identify the first moment of the Nb d-band filling, $\mu_d^{Nb}$, as the primary electronic descriptor controlling the electron–phonon coupling strength and, consequently, the superconducting transition temperature and critical fields. Lattice distortion emerges as a secondary but decisive modifier that weakens coupling and suppresses superconductivity when electronic descriptors are comparable. First-principles calculations confirm that even at fixed stoichiometry, changes in local distortion shift $\mu_d^{Nb}$ relative to the Fermi level, $E_F$, and can significantly modify the superconducting transition temperature, consistent with experimental trends across the alloy series. Our results show that both the electronic and vibrational descriptors that exhibit the strongest correlations with $T_c$ and $H_{c2}$ are Nb-centered, indicating that superconductivity in these chemically complex alloys remains predominantly governed by the local electronic structure and lattice dynamics of Nb. These findings establish a unified design principle for Nb-based superconducting high-entropy alloys in which positioning the Nb d-band center closer to $E_F$ through targeted alloying or strain, while controlling lattice distortion and stiffness, enables simultaneous enhancement of $T_c$ and $H_{C2}$. Elements such as Ti, Zr, Hf, Sc, or Y, with *d*-states closer to the Fermi level, therefore, provide promising pathways for engineering robust superconductors with optimized critical temperature and field performance. More broadly, electronic descriptors such as $\mu_d^{Nb}$, together with lattice distortion and phonon metrics including the average Nb phonon center, $\bar{\omega}_{Nb}$, form a compact and physically interpretable set of variables that enable predictive modeling approaches for reliable estimation of $T_c$ and $\lambda_{e-ph}$ in multicomponent high-entropy alloys.

## Methods
**Sample Preparation**. Rod samples were manufactured using a vacuum arc-melting technique with high-purity elements (99.99 wt % purity), see the detailed experimental procedure[9]. The fabricated rods were sealed in evacuated (10$^{-2}$ torr) quartz tubes with triple-pumped argon prior to heat treatment. Most alloys were heat-treated at 1,200 °C for 3 days, while the NbTaTiVZr alloy was heat-treated at 1,200 °C for 12 hours. All samples were subsequently water-quenched to retain the high-temperature microstructure. The samples were then sectioned to a thickness of less than 1 mm, surface-ground, and flattened on both sides by manual polishing for the resistivity, specific heat, and magnetization measurements. The synchrotron XRD experiments for the homogenized samples were performed using high-energy synchrotron XRD at the beamline 11 ID-C, Advanced Photon Source (APS), Argonne National Laboratory. Theoretical estimation of lattice distortion ($\bar{u}^D$) has been discussed and estimated for NbTa, NbTaTi, NbTaTiV, NbV, and NbTaTiVZr[9]. Equation (1) is used to estimate the lattice distortion ($\bar{u}^D$), where $d_i^{eff}$ is the effective lattice constant of the $i^{th}$ element, and $\bar{d}$ is the average lattice constant of n elements, which are obtained from the atomic size of the incorporated elements related to the crystal structure. $d_i^{eff}$ is estimated by considering the variation of the atomic volume of the $i^{th}$ element in $j^{th}$ element, $\Delta V_{ij}$, across the compositions along with the atomic fractions ($f_j$) in Equation (2).

$$\bar{u}^D = \sqrt{\sum_i^n \left(d_i^{eff} - \bar{d}\right)^2 / n\, d_i^{eff}} \tag{1}$$

$$d_i^{eff} = \sum_j^n f_j \left(1 + \Delta V_{ij}/V_i\right)^{1/3} d_i \tag{2}$$

where, $V_i$ is the atomic volume and $d_i$ is the lattice parameter of the $i^{th}$ element. Theoretically estimated lattice distortion values of alloys have been indicated in Fig. 1d.

**Thermal Characterization.** Measurements were conducted using the Dynacool Physical Property Measurement System (PPMS) from Quantum Design. The PPMS heat capacity option employs relaxation techniques to determine the specific heat of the samples. All Nb-based high-entropy alloy (HEA) samples were diced to dimensions that fit within the mounting platform (approximately 3 mm × 3 mm). Measurements were carried out over the temperature range from 300 K down to 2 K. During each measurement, a known heat pulse was applied to the sample platform, and the resulting temperature response was recorded. The magnitude of the heat pulse raised the sample temperature by approximately 2% above the base temperature. The sample chamber was evacuated to a high vacuum (~ $10^{-5}$ torr) to ensure proper thermal isolation. An addenda measurement was first performed using N-grease, which provides good thermal contact and performs reliably at low temperatures. After obtaining the addenda heat capacity (N-grease + mounting platform), the total heat capacity of the sample assembly (N-grease + mounting platform + sample) was measured. The heat capacity of the sample was then determined by subtracting the addenda contribution from the total measured heat capacity. Data points were acquired once thermal equilibrium was established, and the relaxation data were fit using the PPMS two-tau model. The uncertainties from the two-tau fitting routine were adopted as the primary error estimates.

**Magnetic Characterization.** Magnetic measurements of the Nb-based alloys (Nb, NbTa, NbTaTi, NbTaTiV, NbTaTiVZr) were carried out by measuring the static (DC) magnetic moment as a function of temperature and magnetic field using the Vibrating Sample Magnetometer (VSM) option of the Dynacool PPMS. The system provides a sensitivity of approximately $1.5 \times 10^{-6}$ emu at 300 K. Samples of suitable dimensions were mounted on a quartz rod using Kapton tape to minimize additional diamagnetic contributions. Zero-Field-Cooled (ZFC) and Field-Cooled (FC) magnetization measurements were performed in the temperature range of 2 K to 15 K under an applied magnetic field of 50 Oe (H ∥ ab) for all samples to determine their superconducting critical temperatures. Isothermal magnetization, M(H), was measured below the respective critical temperatures. Prior to each measurement sequence, the system was demagnetized to remove any remanent magnetic field. Subsequently, M(T) measurements were conducted while sweeping the magnetic field from 0 to 9 T, depending on the material. All measurements were performed under high vacuum conditions to ensure thermal and experimental stability. Both M(T) and M(H) scans were carried out using consistent measurement protocols to ensure reliable comparison across samples.

**Resistivity Characterization.** Resistance measurements were carried out using a standard four-probe Kelvin configuration on bar-shaped samples. The samples were mounted on the electrical transport option (ETO) resistivity puck of the PPMS system, and electrical contacts were formed by wire bonding using 1.25-mil Al–Si alloy wire. A constant-current mode was used, and for each sample the excitation current was selected based on its room-temperature resistance to ensure operation within the linear (ohmic) regime, such that the measured resistance remained independent of current and no self-heating occurred. The measured resistance values were converted into resistivity using the known sample geometry, where the resistivity (ρ) was calculated as ρ = R·A/L, with R being the measured resistance, A the sample cross-sectional area, and L the spacing between the voltage contacts (taken as approximately 20% of the total sample length to minimize geometric uncertainty).

**Computational Details.** DFT calculations were performed using *Quantum ESPRESSO* (v7.3)[55,56] with the PBE-GGA exchange–correlation functional[57] and SG15 Optimized Norm-Conserving Vanderbilt (ONCV) pseudopotentials[58,59]. A plane-wave cutoff of 60 Ry and a charge-density cutoff of 400 Ry were used with a 4 × 4 × 4 Monkhorst–Pack $k$-point mesh. The self-consistent-field (SCF) and ionic relaxation thresholds were set to $10^{-10}$ Ry and $10^{-8}$ Ry/Bohr, respectively. All alloys were modeled as 16-atom body-centered-cubic (BCC, 2 × 2 × 2) supercells constructed via the Special Quasi-Random Structure (SQS) approach using *ATAT*[60]. Phonon calculations were performed using *ph.x* on a 2 × 2 × 2 $q$-point grid with a self-consistency threshold of $10^{-15}$ with density functional perturbation theory (DFPT). Electron–phonon coupling and superconducting properties were evaluated using *EPW*[61,62] with dense Brillouin-zone meshes (16 × 16 × 16 $k$-points, 8 × 8 × 8 $q$-points). The Fermi-surface broadening was set to 0.5 eV, the electronic smearing to 0.1 eV, and the phonon smearing to 0.5 meV. The superconducting transition temperature, $T_c$ was estimated using the Allen–Dynes modified McMillan equation with a Coulomb pseudopotential of $\mu^* = 0.13$.

**Acknowledgments**

This research used resources of the Advanced Photon Source, a U.S. Department of Energy (DOE) Office of Science User Facility operated for the DOE Office of Science by Argonne National Laboratory. We thank Dr. M. Adams and Dr. W. Jin for providing access to the characterization facilities at the Alabama Micro/Nano Science and Technology Center (AMNSTC). H.G. acknowledges support from the U.S. Department of Energy, Office of Science, under Grant No. DE-SC0023478. C.L. acknowledges support from the U.S. National Science Foundation, under Grant No. DMREF-2522654. X. Chen acknowledges support from the National Science Foundation under Grant No. 2144328. V.K. and K.K. are supported in part by the NASA Alabama EPSCoR Research Seed Grant program under Grant No. AL-80NSSC22M0050. V. K. acknowledges financial support from the Alabama Graduate Research Scholars Program (GRSP), funded through the Alabama Commission on Higher Education and administered by the Alabama EPSCoR.

**Contributions**

F.K. and C.L. conceived the idea. F.K. supervised the project and led the manuscript preparation with input from all authors. M.S.H.B. and H.G. performed electrical resistivity, specific heat, and magnetic characterization experiments under the supervision of F.K and M.A. V.K. carried out the calculations under the supervision of K.K. D.P.N. synthesized the materials and performed XRD and SEM characterization. Y.X. and S.L. conducted independent electrical resistivity measurements under the supervision of X.C. S.E.P. performed EBSD characterization under the supervision of B.C.P. All authors contributed to data analysis, discussion of the results, and manuscript revision.